\documentclass[useAMS,usenatbib,onecolumn]{mn2e}
\usepackage{graphicx}
\renewcommand{\v}[1]{\ensuremath{\mathbf{\stackrel{\rightarrow}{#1}}}}
\title[Generalized-SCIDAR vertical wind  profiles]{Processing of turbulence-layer wind speed with
 Generalized SCIDAR through wavelet analysis}

\author[Garc\'{\i}a-Lorenzo et al.]{B. Garc\'{\i}a-Lorenzo$^{1}$\thanks{E-mail:
bgarcia@iac.es} and J. J. Fuensalida$^{1}$ \\
1 Instituto de Astrof\'{\i}sica de Canarias, C/Via Lactea S/N, 38305-La Laguna, Tenerife, Spain}

\begin{document}

\date{Accepted ..... Received .....; in original form .....}

\pagerange{\pageref{firstpage}--\pageref{lastpage}} \pubyear{2006}

\maketitle

\label{firstpage}

\begin{abstract} 
We describe a new method involving wavelet transforms for deriving the
 wind velocity associated with atmospheric turbulence layers from
 Generalized SCIDAR measurements. The algorithm analyses the cross-correlation
of a series of scintillation patterns separated by lapses of $\Delta t$,
 2$\Delta t$, 3$\Delta t$, 4$\Delta t$ and 5$\Delta t$ using wavelet
 transforms. Wavelet analysis provides the position, direction and
 altitude of the different turbulence layers detected in each 
 cross-correlation. The comparison and consistency of the turbulent layer
 displacements in consecutive cross-correlations allow the determination of their
 velocities and avoid misidentifications associated with noise and/or overlapping
 layers. To validate the algorithm, we have compared the  velocity of
 turbulence layers derived on four  nights with the wind vertical profile
 provided by balloon measurements. The  software is fully automated and
is able to analyse huge amounts of Generalized SCIDAR measurements.
\end{abstract}


\section{Introduction}
Adaptive optics in astronomy is a poweful technique that compensates for
 the distortions introduced by the atmosphere on the quality and resolution
 of images taken by ground-based telescopes. The excellent image quality
 requirements of the new generation of large and extremely large telescopes
 demands a proper knowledge of atmospheric turbulence in order to design 
efficient adaptive optics systems. Knowledge of the spatial and temporal behaviour of  
atmospheric turbulence at an astronomical site is crucial for optimizing the
efficiency and operation of adaptive optics systems working with several
 deformable mirrors (multiconjugate adaptive optics).
 The SCIntillation Detection And Ranging technique (SCIDAR) is the
 most contrasted and efficient remote sensing technique for obtaining the optical
 vertical structure of  atmospheric turbulence. The SCIDAR technique, its
 implementation and results have been extensively explained in several papers 
(Vernin \& Roddier 1973; Rocca, Roddier \& Vernin 1974;
 Fuchs, Tallon \& Vernin 1994; Avila, Vernin \& Masciadri 1997;
 Kluckers et al.\ 1998; Johnston et al.\ 2002).

In the last few years, several campaigns for atmospheric turbulence
 characterization have been carried out at different astronomical sites, many
of them using the generalized SCIDAR (G-SCIDAR hereafter) technique (Avila, Vernin \& Cuevas 1998;
 Avila et al.\ 2003; Kluckers et al.\ 1998; Vernin et al.\ 2000; McKenna et al.\
 2003; Fuensalida et al.\ 2004a,b; Avila et al. 2006). The refractive index 
structure constant  profiles, $C_{N}^{2}$(h), are derived
 from SCIDAR observations through automatic programs using
 inversion methods (Vernin 1992; Kluckers et al.\ 1998; Prieur, Daigne
 \& Avila 2001; Johnston et al.\ 2002). The velocities of the turbulent layers
are obtained through interactive programs (Kluckers et al.\ 1998;
 Avila et al.\ 2001, 2003; Vernin et al.\ 2000) based on the CLEAN method, 
(Prieur et al.\ 2001) and the first iterative and potentially automated 
algorithm also based  on the CLEAN procedure has been recently published 
(Prieur et al.\ 2004). We present here an alternative method for deriving the
 wind velocity of turbulence layers
 from G-SCIDAR measurements based on wavelet analysis. We have
 developed  fully automatic software that is very
 suitable for analysing huge amounts of SCIDAR data based on this new method.

 In Section \S\ref{desc} we briefly describe the G-SCIDAR technique. A
 brief introduction to  wavelet analysis applied to the problem of
 determining wind turbulence profiles is presented in Section \S\ref{wa}. The input
 data for the algorithm are shown in Section \S\ref{we} and we describe the algorithm
 in Section \S\ref{result}. The velocity of
 turbulence layers derived from our wavelet-based method are compared to wind
 vertical profiles measured using balloons (Section 5). In this paper we 
 present only the developed wavelet-based method and  results derived from a few
 G-SCIDAR observations compared to balloon measurements to test the
proposed algorithm. We are already preparing a paper with the statistical 
results of the wind velocity of turbulent layers above the Canary Islands 
astronomical observatories from long-term SCIDAR measurements.

\section{Brief description of the G-SCIDAR technique}
\label{desc}

Classical techniques (Vernin \& Roddier 1973; Rocca, Roddier \& Vernin 1974) and
 G-SCIDAR (Fuchs, Tallon \& Vernin 1994;
 Avila, Vernin \& Masciadri 1997; Kluckers et al.\ 1998) analyse
 the scintillation patterns produced at the
 telescope pupil by the light coming from the  stars in a binary system.
 Turbulence profiles
 as a function of height, $C_{N}^{2}(h)$, are derived
 through the inversion of the average normalized autocovariance of a large
 number of scintillation patterns (see Fuchs et al.\ 1994 and references
 therein for a detailed description of  G-SCIDAR theory).

\begin{figure}
\centering
\includegraphics[scale=0.5]{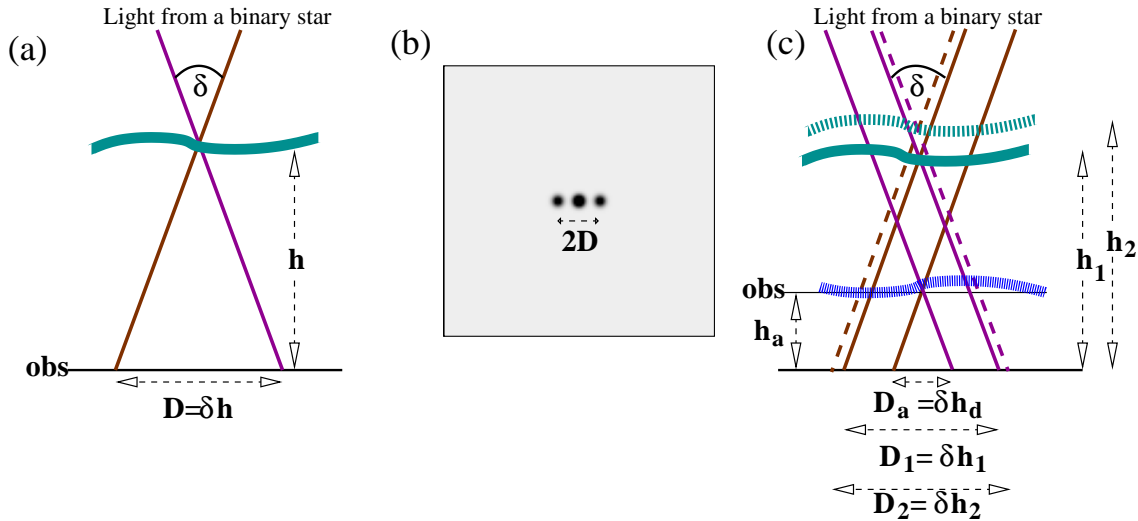}
 \caption{(a) Principle of the SCIDAR technique considering a single turbulence
 layer. {$\delta$} corresponds to the angular separation between the stars
 in the binary system. {\itshape h} is the altitude of the turbulence layer over
 the observation plane. {\itshape D} corresponds to the distance between the lateral
 peaks of the normalized autocorrelation. (b) The average normalized
 autocorrelation of a series of scintillation
 patterns at the telescope pupil produced by a single turbulent layer on the
 light coming from a binary star. (c) The principle of G-SCIDAR 
considering multiple turbulence layers. {$h_{1}$} and {$h_{2}$} indicate the altitude of two turbulence layer over the analysis plane. 
The {\itshape obs} line  represents observatory level. {$h_{a}$} corresponds to the position of the analysis plane 
below observatory level. {$D_{1}$}, {$D_{2}$} and {$D_{a}$} indicate the distance between the lateral
 peaks of the normalized autocorrelation produced for the different turbulence layers considered. }
\label{figure2}
\end{figure}

\begin{figure}
\centering
\includegraphics[scale=0.5]{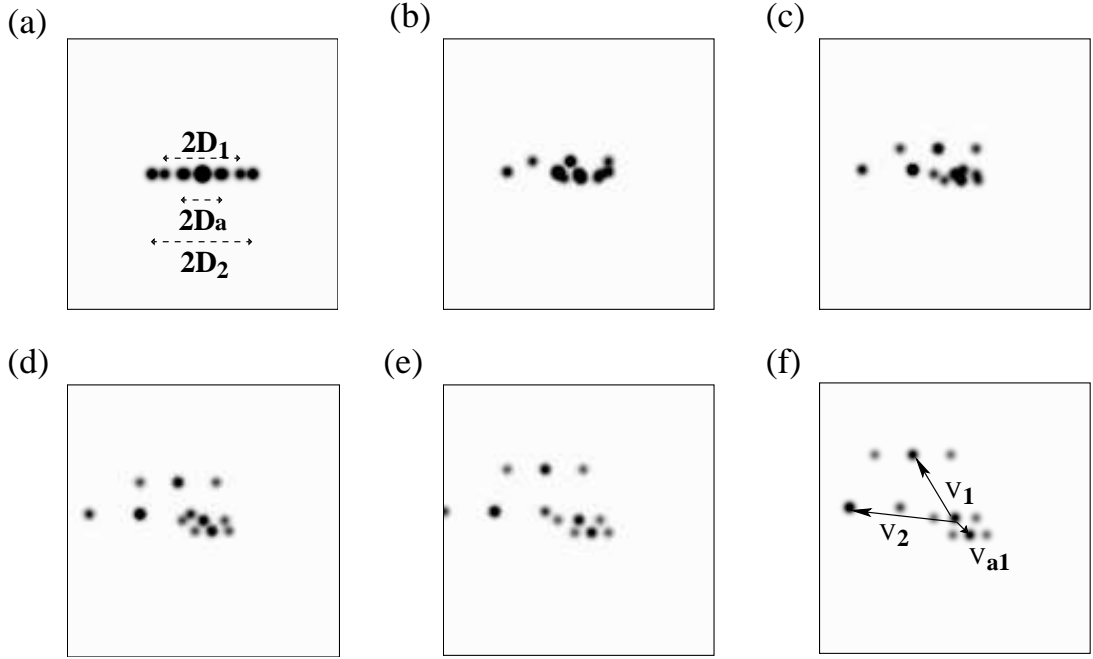}
 \caption{Maps in this figure are only simulations to illustrate the problem and 
 do not correspond to real data.(a) Simulation of the average normalized
 autocorrelation of a series of scintillation
 patterns at the telescope pupil produced by three turbulent layers on the
 light coming from a binary star. {$D_{1}$}, {$D_{2}$} and {$D_{a}$} indicate the distance between the lateral
 peaks of the normalized autocorrelation produced for the different turbulence layers considered as in Figure \ref{figure2}c. 
 (b) Simulated cross-correlation of scintillation images
 separated by a lapse $\Delta t$. (c) Simulated cross-correlation of scintillation images
 separated by a lapse 2$\Delta t$. (d) Simulated cross-correlation of scintillation images
 separated by a lapse 3$\Delta t$. (e) Simulated cross-correlation of scintillation images
 separated by a lapse 4$\Delta t$. (f) Simulated cross-correlation of scintillation images
 separated by a lapse 5$\Delta t$. $V_{1}$, and $V_{2}$ are the wind speeds of
 two turbulence layers over the observing site, while $V_{a1}$ corresponds to
 the speed of a turbulent layer close to the altitude of the observing site.
    }
\label{figure3}
\end{figure}

 While  classical SCIDAR was insensitive to turbulence at low altitudes, with 
 G-SCIDAR  low altitude atmospheric turbulence can be separated
 from the dome contribution (Fuchs, Tallon \& Vernin 1994). In  G-SCIDAR, the plane of observations
 lies a few kilometres below the telescope pupil, while the telescope pupil
 is the analysis plane in the classical SCIDAR.

 For a single turbulent layer at an altitude {\itshape h} above the ground (Fig. \ref{figure2}a), 
 the average normalized autocorrelation of several scintillation patterns from SCIDAR
 measurements of a binary system will result in a frame with three peaks
 (Fig. \ref{figure2}b). The distance {\itshape D} between the lateral and central peaks is
 related to the angular separation ($\delta$) of the two stars forming the binary system
 and the altitude of the turbulence layer ({\itshape h}), as is
 illustrated in Figure \ref{figure2}a.
When the atmosphere presents several turbulent layers (Fig. \ref{figure2}c) the result
 from the analysis of G-SCIDAR measurements will be a frame with
 several
 lateral peaks (Fig. \ref{figure3}a). In this map, the central peak includes the
 contribution of the different turbulent layers. Lateral peaks equidistant from the central peak indicate the intensity and altitude of a particular
 turbulent layer. $C_{N}^{2}(h)$ is basically derived through the
 inversion of the cut connecting the peaks (Vernin 1992;
 Kluckers et al.\ 1998; Prieur, Daigne \& Avila 2001; Johnston et al.\ 2002).

 Wind velocity of turbulence layers can be derived from the average
 normalized cross-correlation of a series of scintillation patterns relative
 to a reference pattern (Avila et al.\ 2001). If we consider a single turbulent layer with a wind
 velocity \v{V} , the average normalized cross-correlation of two scintillation
 patterns taken at times separated by $\Delta t$ will result in a map with
a triplet placed at \v{V}$\Delta t$ from the central peak of the corresponding 
average normalized autocorrelation. In the most realistic case of
 several turbulent layers moving at different wind velocities, the resulting frame
 will show several triplets located at different positions according to the
 wind vector of each layer (Fig. \ref{figure3}b). When we take frames separated
by $n\Delta t$, the triplets will placed at $n$\v{V}$\Delta t$ from the central peak (0,0).
Figures 3c--f show the average normalized cross-correlation of two
 scintillation patterns taken at 2$\Delta t$, 3$\Delta t$, 4$\Delta t$ and
 5$\Delta t$, respectively. In these simulated cross-correlations, layers are
 well resolved for scintillation patterns taken after 3$\Delta t$, which is not
 far from the real case. The altitude of the different turbulent layers is
 determined by the distance of the lateral to the central peak of each triplet,
 as in the autocorrelation. The distance and relative position of triplets
with  respect to the centre of the frame determine the wind speed of each
 turbulence layer. Information on the velocity of relatively faint
 turbulent layers can eventually be missed in the cross-correlation owing to
 temporal decorrelation of the scintillation and/or eventual fluctuations
 of \v{V} during the integration time (Avila et al.\ 2001).

It is important to note that the SCIDAR technique  provides wind speed 
measurements only where a turbulence layer is detected.

\section{Wavelet-based algorithm for deriving turbulence layer wind speed}

\subsection{Wavelet-based transform application}
\label{wa}

Wavelets are mathematical functions that decompose data into different frequency components allowing
each component to be studied separately.  The basic idea behind wavelet analysis is to 
take a mother wavelet ($M(X)$), translate and dilate it ($M_{a,b}(X)$), integrate it product with the signal $(F(X))$ and study the coefficients in  {\it wavelet space},
 spanned by translation {\it b} and dilation {\it a}. The set
 of functions $M_{a,b}(X)$ are given by

\begin{equation}
\centering M_{a,b}(X) = \frac{1}{a^{1/2}}M(\frac{X-b}{a}),
\end{equation}
where {\it a} and {\it b} are real,  {\it a} being greater than 0. Wavelet
 analysis is based on the integration of the $F(X)$ products with the set of functions
 $M_{a,b}(X)$ :
\begin{equation}
T_{M}({\it b,a}) = \frac{1}{a^{1/2}}\int M^* (\frac{X-b}{a}) F(X) dX,
\label{WT}
\end{equation}
where {\it b} is the translation parameter and corresponds to position or time when
 the data is spatial or temporal, respectively.  The dilation parameter {\it a}
 corresponds to scale length or temporal period. M$^{*}$ indicates the complex conjugate of M. The wavelet transform
 (Equation \ref{WT}) expands a one-dimesional function, $F(X)$, into a
 two-dimensional parameter space, the {\it wavelet space} ({\it b}, {\it a})
 giving a local measure of the relative amplitude of activity at scale {\it a}
 and at space/time {\it b}. Wavelet analysis has been applied in a 
wide variety of scientific fields, including atmospheric sciences, engineering,
 financial analysis, geophysics, image analysis, medical science, turbulence, 
etc. A practical introduction and description of wavelet analysis
can be found, for example, in Torrence \& Compo (1998) and Mayers, Kelly \& O'Brien 
(1993).

\begin{figure}
\centering
\includegraphics[scale=0.45]{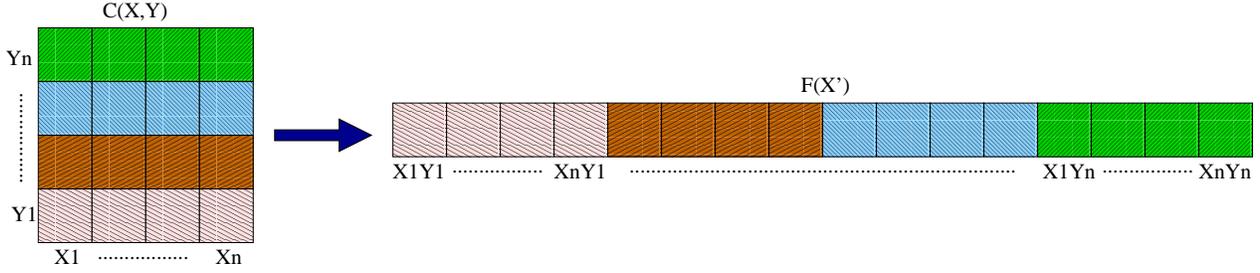}
 \caption{Transformation of each cross-correlation $C(X,Y)$ derived from SCIDAR observations into a one-dimensional 
 function $F(X')$ to apply the 1D wavelet algorithm.}
\label{2Dto1D}
\end{figure}

In order to derive the velocity of the turbulence layers from SCIDAR
 observations using wavelets, we have transformed each cross-correlation
 ($C(X,Y)$) into a one-dimensional function ($F(X')$) placing the different 
lines $Y$ consecutively (see Fig. \ref{2Dto1D}). Therefore, $F(X')$ is a one-dimensional function of $X\times Y$ 
elements where each set of $X$ elements corresponds to a $Y$ line in the $C(X,Y)$. Hence, $F(X')$ is a function in  
spatial space, where each element corresponds to a well-define position in the $C(X,Y)$. In the {\it wavelet space}, 
the translation parameter {\it b} can be identified with the position $X'$ and hence to the position in the 
cross-correlation
 $C(X,Y)$. Therefore, {\it b} gives information about wind speed (velocity and direction). The dilation parameter {\it a} 
 is related to the distance between the lateral peaks to the central peak of each turbulent layer. Hence, {\it a} can be 
 identified to the altitude of the different turbulent layers.

\begin{figure}
\centering
\includegraphics[scale=0.95]{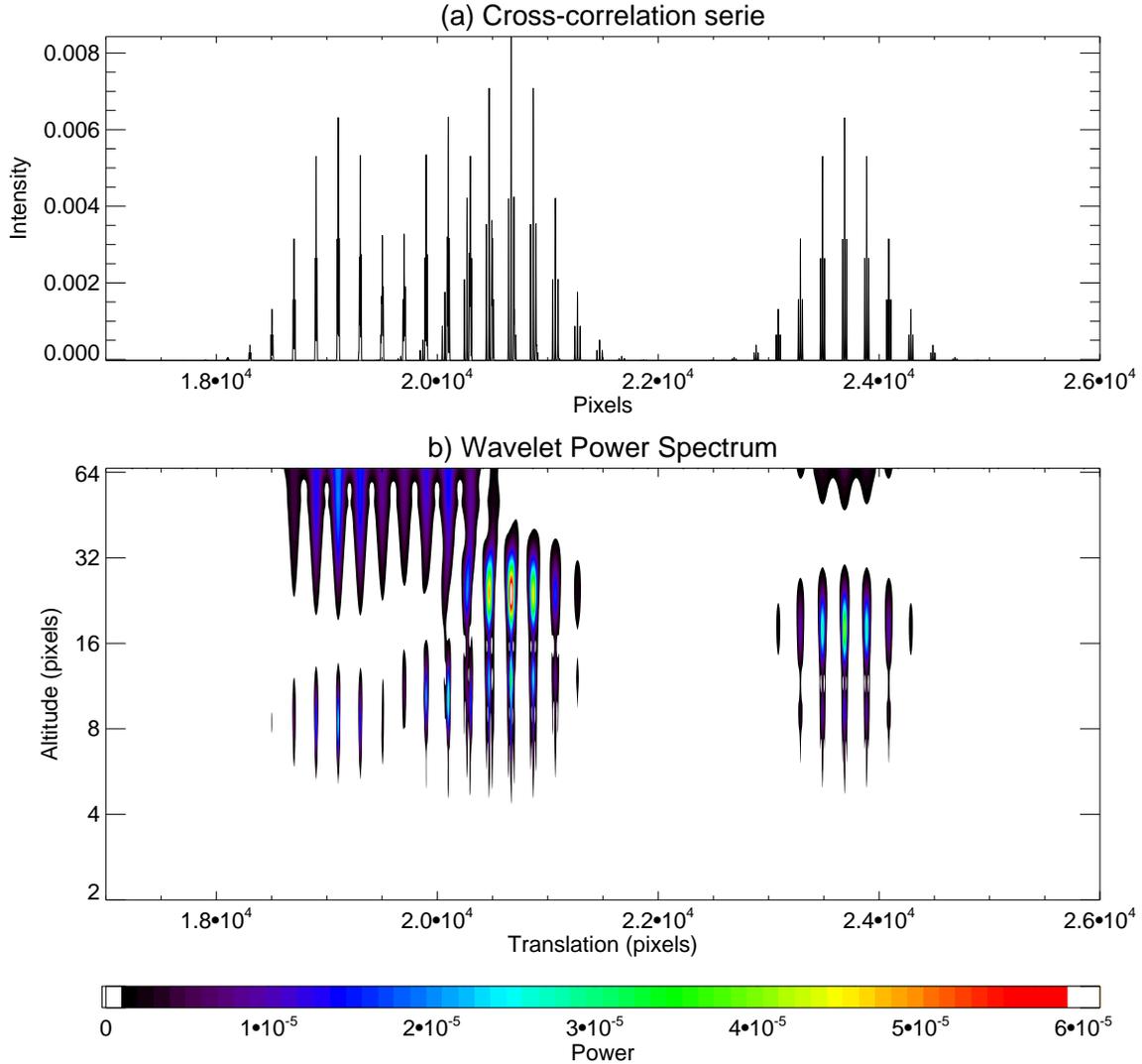}
 \caption{(a) The 1D cross-correlation series derived from the cross-correlation
 of Fig. \ref{figure3}d, as explained in Figure \ref{2Dto1D}. (b) The
 wavelet power spectrum derived through the convolution of the 1D
 cross-correlation series with a set of Morlet wavelets (a dilated and translated Morlet function). 
The horizontal axis of this plot corresponds to a 
well-defined position in the cross-correlation, and the vertical axis
 indicates the separation of the lateral peaks of each detected triplet in
 the cross-correlation. The brightest peaks in this 2D map corresponds
 to position of the central peak of each triplet. }
\label{ejemplo_wavelet}
\end{figure}

 For the analysis of  $F(X')$, we have selected a commonly used mother wavelet, the Morlet function. The Morlet 
 wavelet consists of a plane wave modulated by a Gaussian,

\begin{equation}
M(x) \approx e^{i\omega_{0}x} e^{-x^{2}/2},
\end{equation}
where $\omega_{0}$ is a non-dimensional scale length (space) or frequency (time) that we take as 
6\footnote{The Morlet function is only admissible as a wavelet if some certain correction terms are 
added to satisfy the zero mean condition imposed for a wavelet. In practice, when $\omega_{0}=6$, 
the correction terms are very small and of the same order as computer round-off errors (see Farge 1992).} to 
satisfy the condition that the wavelet function  have zero mean and be located in both space/time and scale/frequency 
space (Farge 1992).
 Considering the form of the Morlet wavelet, equation (2) is only the convolution of the signal, F(X') with a set of Morlet functions. The result of the convolution of $F(X')$ with Morlet wavelets (a sequence of
 translated and dilated Morlet functions) is a 2D function in wavelet space
 ({\it b},{\it a}) with the brightest peaks indicating the wind speed ({\it b}
 component) and the turbulent layer altitude ({\it a} component). Figure 
\ref{ejemplo_wavelet} shows an example of the 2D function in the {\it wavelet space derived} from the 
convolution of the simulated cross-correlation frame in Figure \ref{figure3}d transformed
 to a 1D function (as in Figure \ref{2Dto1D}) and a series of Morlet functions.
The wavelet power spectrum
 presents several bright knots, the brighter ones corresponding to each of the turbulent layers. 
Fainter knots placed at the same translation position as the brighter ones correspond to harmonic frequencies 
(see also Fig. \ref{validation}). In Section \S\ref{we} we explain the way
we select or reject a peak in the wavelet power spectrum to determine the
 wind speed of the turbulence layers. 

\subsection{Wavelet-based algorithm for wind velocity measurements}
\label{we}

The algorithm proposed in this paper is based on wavelet transforms. As we
 explained in Section \S\ref{wa}, the peaks of the wavelet power spectrum derived
 from the convolution of the 1D cross-correlation with a set of Morlet
 functions provide information about the location of the triplets in
 the input data, as well as the separation of the lateral peaks. 
The basic idea of the method is to extract the velocity of the turbulent
 layers by the consistencies of its displacements in consecutive wavelet power
 spectra obtained from the cross-correlation of scintillation patterns taken at intervals $\Delta t$, 
2$\Delta t$, 3$\Delta t$, 4$\Delta t$ and 5$\Delta t$. As we have five
 cross-correlations, we have five different 1D cross-correlation inputs
for the wavelet analysis. Therefore, we get five wavelet power spectra.
The algorithm has already been tested with simulated data (Garc\'{\i}a-Lorenzo \& 
Fuensalida 2006).

The first step in the processing of the data is  noise filtering in the
 cross-correlation frames. 
 Different sources contribute to the
 noisy background: noise from the detector, electronic noise, and noise
 associated with the finite dimensions of the images. We first
 identify the pixels in the cross-correlations (and auto-correlation) with
 values smaller than zero and change their value to zero. From the
 autocorrelation, we eliminate the central pixels, where the triplets
 corresponding to the turbulence profiles appear (Fig. \ref{figure3}a), and
 we calculate the background median of the pixels larger than zero ({\it BM}
 hereafter) and their standard deviation ({\it ST} hereafter). {\it BM} and 
{\it ST} are taken as references for the noise filter. The noisy background
 of the cross-correlations is evaluated as the pixels in these maps
 with values smaller than {\it BM} plus a factor of {\it ST}. A default
 value of 3 is adopted for the automatic processing of the data, although this 
 factor can also
be  selected by the user when starting the analysis of the data
 with the developed
 software. If a factor two is selected, we call a {\it sigma2} filter for the
analysis. For a factor four, we call a {\it sigma4} filter. Those pixels 
satisfying the noise filtering conditions are taken as
 zero, and only pixels with values larger than the noise filter conditions are
 considered as real signal. 

\begin{figure}
\centering
\includegraphics[scale=0.7]{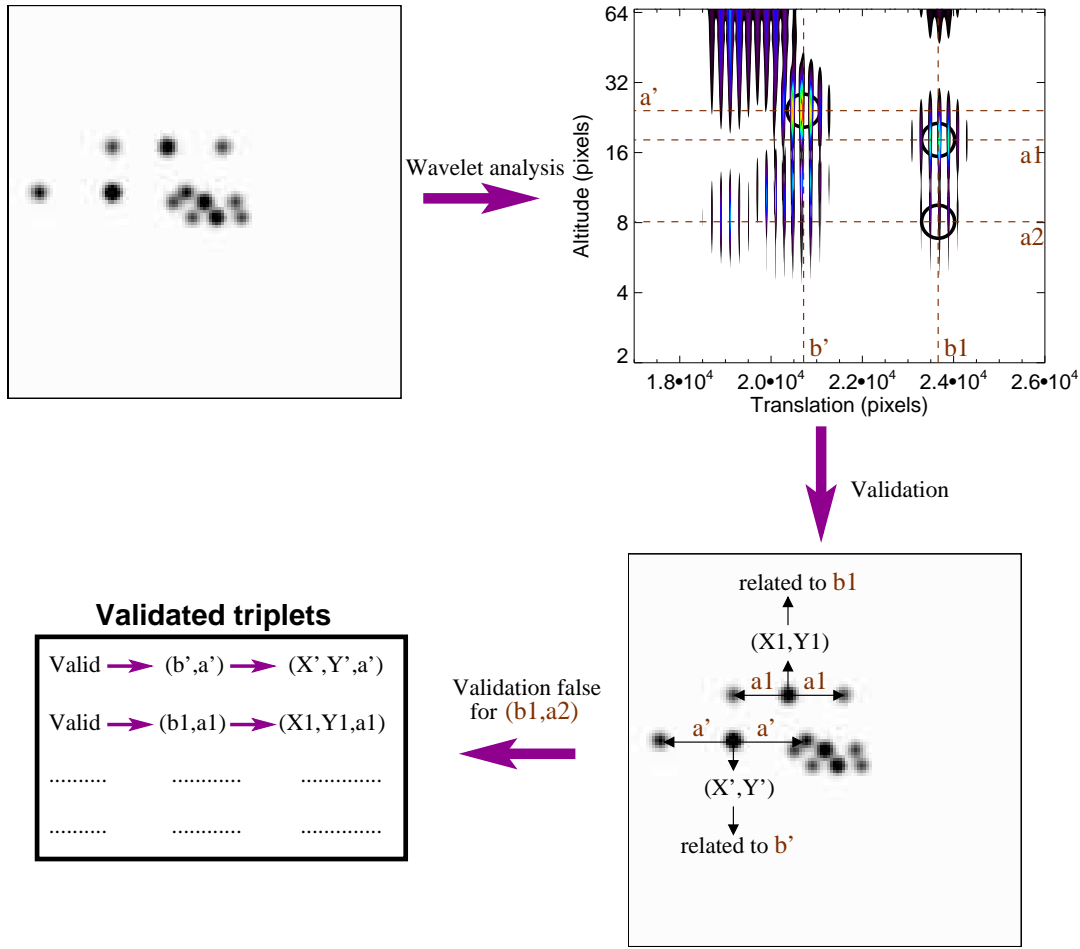}
 \caption{Schematic view of the procedure for validating triplets located in the 
wavelet power spectrum from the wavelet analysis of a cross-correlation.
 In the top left panel, we present the input data, the cross-correlation at
 $n\Delta t$. The top-right panel corresponds to the wavelet power spectrum 
derived from the wavelet analysis. The filled circles mark the
 selected points to explain the method for validating the triplets. The 
bottom-right panel indicates the points giving a positive validation. The file 
with the validated triplets, their corresponding position in the
 cross-correlation and the separation between the lateral peaks is
 shown in the bottom-left panel.}
\label{validation}
\end{figure}

 The wavelet algorithm starts analysing the cross-correlation corresponding to
 5$\Delta t$, which we refer to hereafter as cc5. The
peaks located in the wavelet power spectrum are caracterized by two
 coordinates. The horizontal coordinate gives information about the location
 of the triplet in cc5 (the velocity vector), and the vertical
 coordinate corresponds to the separation of the lateral peaks of the triplets
(the height of the turbulent layer). In order to discard
 any harmonic frequency or any peak associated with noise, we compare the
 derived coordinates with 5cc. For each peak detected in the
 wavelet power spectrum, we translate its coordinates to cc5 and
 check that, at the calculated position, there exists a triplet with the lateral
 peaks separated by as much as the vertical coordinate indicates. If the
 validation is positive, the coordinates of that peak are saved in an
 intermediate file. Those peaks of the wavelet power spectrum with a negative
 validation are considered as harmonics or noise and are deleted from the rest
 of the analysis. Figure \ref{validation} presents an outline of the procedure
over the cross-correlation shown in Figure \ref{figure3}d. Sometimes,
 a few peaks that are not associated with any real triplet but with noise are not
 eliminated in this first step of the algorithm. 

We follow the same procedure for the cross-correlations at 4$\Delta t$ (cc4
 hereafter) and 3$\Delta t$ (cc3 hereafter). Once we have
 proccesed cc5, cc4, and cc3, we translate to 3$\Delta t$ the validated
 triplets for each cross-correlation using the relationship: $Velocity=Space/Time$. In this way, any triplet
 located at coordinates ($X_{n}$,$Y_{n}$) at $n\Delta t$ will be placed at the
 position:

\begin{equation}
(X_3,Y_3) = (X_n,Y_n) \frac{3\Delta t}{n\Delta t}.
\end{equation}

\begin{figure}
\centering
\includegraphics[scale=0.65]{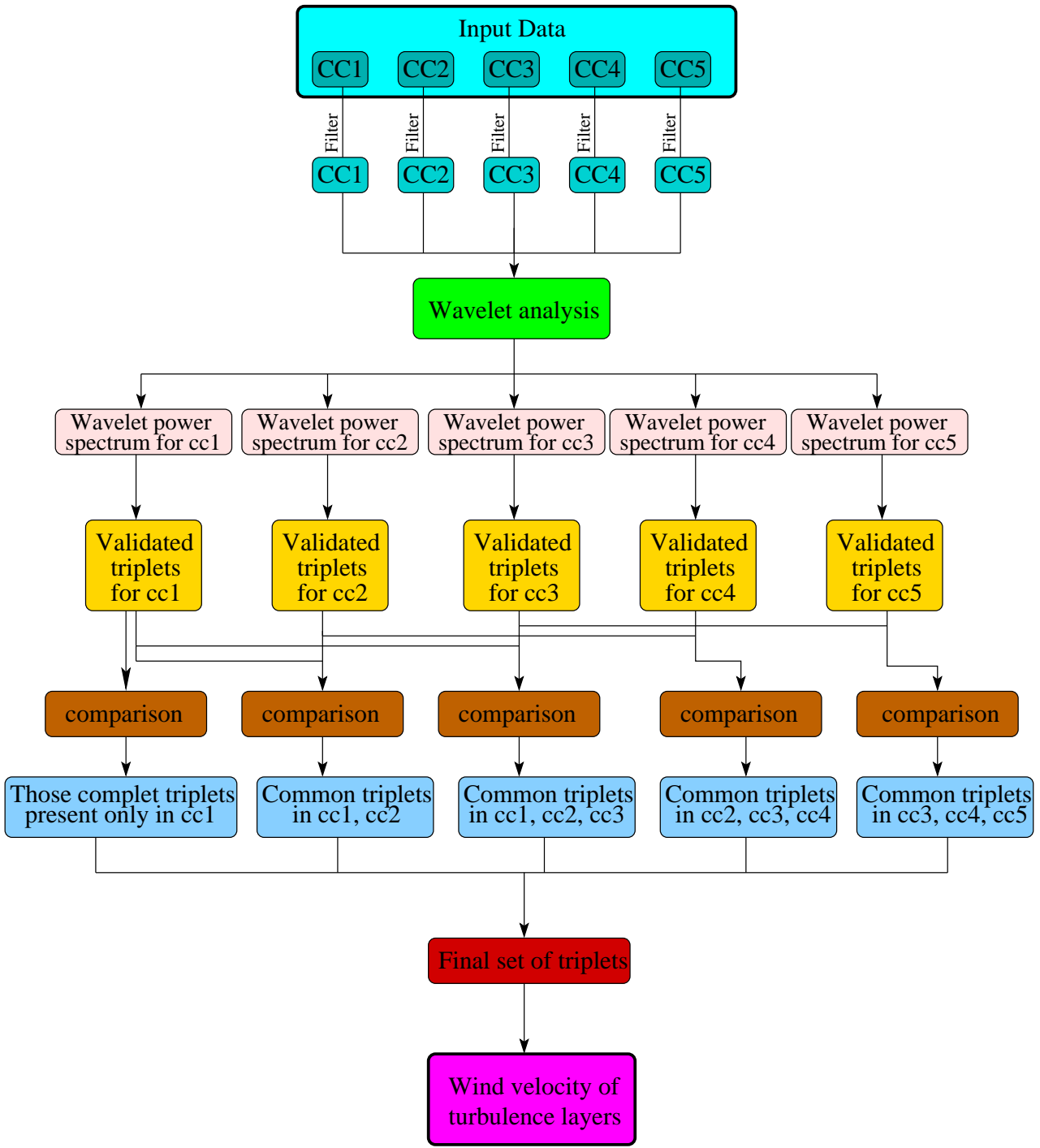}
 \caption{Schematic view of the algorithm for computing wind velocities of turbulence layers for SCIDAR observations (see 
 Section \S\ref{we}).}
\label{algorithm}
\end{figure}

We then compare the three sets of points at the same temporal position
 3$\Delta t$. Those points at the same coordinates and showing the same distance between
 lateral peaks are identified as true triplets. To compare
 coordinates, we allow an uncertainty equal to the full width half maximum of
the central peak of the triplets. It is clear that any triplet in cc5 must be at cc4 and cc3,
 although triplets present at cc4 and/or cc3 corresponding to faster layers
could be missing in cc5. We continue the analysis in the same way,
 now comparing cc4, cc3 and the cross-correlation corresponding to 2$\Delta t$
 (hereafter cc2). Triplets at cc4 must be in cc3 and cc2, but the triplet
 in cc3 and cc2 could be absent in cc4. We compare cc3, cc2 and the
 cross-correlation at $\Delta t$ (hereafter cc1), and finally, we compare the
 results for cc2 and cc1. The triplet in cc2 must be in cc1, although these tend
 to be blended into one another. The coordinates of the final validated 
triplets are transformed to the velocity
 and altitude using the determined parameters of the observational
 setup ($\Delta t$,the conjugated plane, observatory altitude, etc.). 
 To summarize,
 Figure \ref{algorithm} shows a schematic view of the algorithm.  We are able
 to identify high velocity 
turbulent layers if the central and one of the lateral peaks are present in
 cc1 at least. Only those triplets
 associated with turbulence layers moving as fast as they have disappeared in 
 cc1 are undetected in our procedure.
 These kinds of layers should be (a) very fast and in low-altitude or
  (b) high-altitude
 layers of moderate velocity. For our system (see section \S4.1 for 
details), in a configuration with the 
conjugate plane at four kilometres and a binary star with an angular
 separation of 9.6 arcsec, a low-altitude turbulence layer moving faster
 than 45 m/s or a high altitude layer (over 18\,000 m) moving faster
 than 29 m/s
 will be lost from the analysis depending on the direction. Low altitude
 layers faster than 45 m/s
are associated with bad weather conditions, when the telescope is closed due
 to
 wind limitions on telescope operation. For the second case, when a 
high-altitude layer is moving at a moderate velocity the proposed algorithm
 will miss the layer. Indeed, the instrument configuration
 (see Section \S4.1) will miss the velocity information on such turbulence
 layer and only increasing the sample frequency will retain such information.
 High altitude turbulence layers moving faster than 30 m/s have been already
 reported on San
 Pedro M\'artir. However, the statistical study of wind vertical profiles
at different astronomical sites (Garc\'{\i}a-Lorenzo et al. 2005) indicates
that such fast winds do not occur very often at high altitudes. Moreover,
the turbulence above the Canary Islands observatories tend to be concentrated 
in low-altitude layers (Fuensalida et al.\ 2004). Therefore, the second case
 should not be a habitual situation in our data. 

A few programs have been already developed for the
 determination of wind profiles from G-SCIDAR observations
 (Kluckers et al.\ 1998; Avila et al.\ 2001, 2003; Vernin et al.\ 2000; Prieur
 et al.\ 2004). The advantage of our procedure is not only the new algorithm, 
but also that we make use of five cross-correlations instead of two, as
with previous programs and therefore have a higher resolving capacity of slow
 layers. The processing of more maps represents a higher
 computer dedication, but it minimize several common problems. 
The
 detection of dome seeing (layers with $V=0$ in the cross-correlations)
 is easier when using five cross-correlations, which is an advantage of the developed procedure over previous works (Kluckers et al.\ 1998; Avila et al.\ 2001, 2003; Vernin et al.\ 2000; Prieur
 et al.\ 2004). For example, in Fig.\ 3, 
the layer corresponding to dome seeing (the central triplet with $V=0$) is blended
 with other turbulence layers in the cross-correlation at $\Delta t$ (Fig.\ 
 \ref{figure3}b) and 2$\Delta t$ (Fig.\ \ref{figure3}c), while it appears resolved 
from the cross-correlation at 3$\Delta t$ (Fig.\ \ref{figure3}d). Therefore, for
 cross-correlation of scintillation patterns with lapses of 5$\Delta t$, only
 low-altitude turbulence layers with velocities smaller than twice the
 velocity resolution of the system can be confused with dome seeing.

In Section \S\ref{result} we present some examples and explanations to
 illustrate the automatic program.

\section{Wind profile results from the wavelet-based algorithm}
\label{result}

In this section, we present a few examples to illustrate the data processing with the proposed 
algorithm on real G-SCIDAR observations. We compare the derived velocity of the turbulence 
layers with the vertical wind profile measured with balloons.

\subsection{The Data}
\label{data}

The data analysed in this paper were obtained on 2003 July 23, August 5--6 and
 August 31 at the Observatorio del Teide (OT) on the island of Tenerife
 (Spain). The 1.5 m Carlos S\'anchez Telescope (1.5m-TCS) was used in
 combination with the G-SCIDAR instrument developed at the Instituto
 de Astrof\'{\i}sica de Canarias (IAC). Each detector pixel covers a square
  of side 2.81 cm on the 1.5m-TCS pupil. We used 1000 images of
 scintillation patterns to derive the average normalized autocovariance and
 five frames
 that are measurements of the 2D spatiotemporal cross-correlation functions
 separated by a lapse of $\Delta t$ equal to 26 ms, 52 ms, 78 ms, 104 ms and
 130 ms. These sets of frames are similar to those maps in Fig.\ \ref{figure3}
 and they constitute the experimental data input to the wavelet-based
 algorithm in order to determine the wind velocity of the turbulence layers.

For comparison and validation of wind speed derived from the wavelet-based
 algorithm, we have used wind vertical profiles measurements from the
 closest radiosonde station located $\sim$13 km from the OT. The Centro
 Meteorol\'ogico Territorial de Canarias Occidental of the Spanish
 Instituto Nacional de Meteorolog\'{\i}a (INM: http://www.inm.es) operates this
 radiosonde station, which is part of the NOAA network
 (station 60018). Radiosondes are launched twice daily (at 12Z and 00Z)
 from an altitude of 105 m and reaches an altitude of around 30 km.
 We have selected the radiosonde data obtained at 00Z for the nights of our
 SCIDAR observations.

\subsection{Examples of Processing}

\begin{figure}
\centering
\includegraphics[scale=0.4]{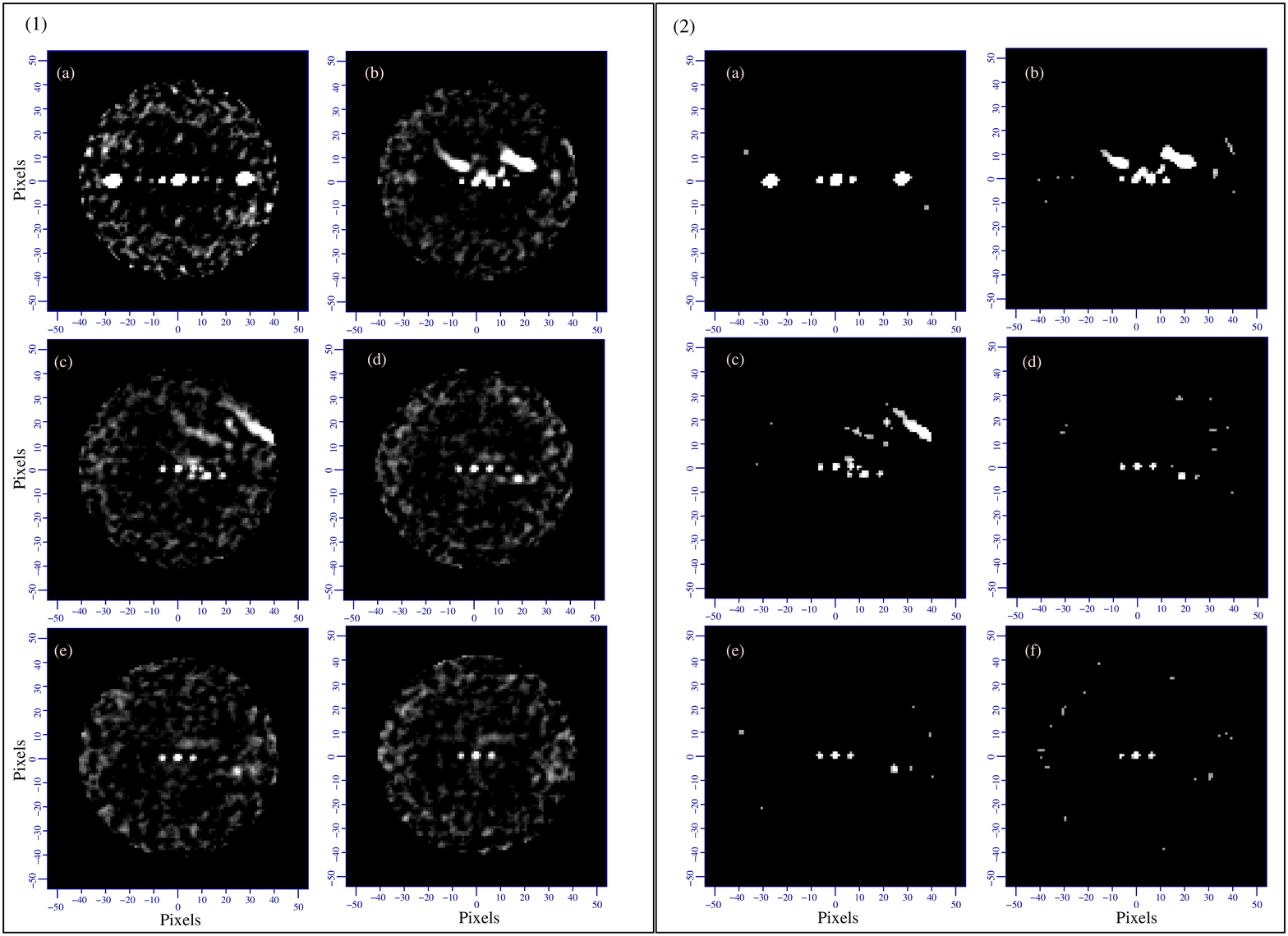}
 \caption{(a) The average normalized autocorrelation of 1000 scintillation
 patterns observed at the TCS pupil produced by turbulent layers on the light
 coming from the binary star BS 7948 on the night of August 31 at around 23 UT. (b) Cross-correlation of
 the same series of scintillation images as (a) but separated by a lapse of
 $\Delta t$ = 26 ms. (c) Cross-correlation of the same series of scintillation images as (a) but
 separated by a lapse of 2$\Delta t$ = 52 ms. (d) Cross-correlation of the same series of
 scintillation images as (a) but separated by a lapse of 3$\Delta t$ = 78 ms. 
 (e) Cross-correlation of the same series of scintillation images as (a) 
but separated by a lapse of 4$\Delta t$ = 104 ms. (f) Cross-correlation of the same
 series of scintillation images as (a) but separated by a lapse of 5$\Delta t$ = 130
 ms. The lefthand plots (1) correspond to drawn frames and the righthand plots (2) are maps after 
the noise filter {\it sigma4} were applied.}
\label{proc1}
\end{figure}

The developed software is able to
 evaluate the wind velocity of turbulent layers for the data of a full night
 or only the wind velocity associate with a particular G-SCIDAR measurement. The
 first example of processing corresponds to the latter case. We have selected a
{\it sigma4} filter for the noise filtering (see Section \S\ref{we}).
 Figure \ref{proc1} shows the autocorrelation (Fig.\ \ref{proc1}a) and the five
 cross-correlations (Fig.\ \ref{proc1}b, \ref{proc1}c, \ref{proc1}d, \ref{proc1}e  and \ref{proc1}f) 
frames corresponding to that particular measurement before (Fig.\ \ref{proc1}(1)) and after 
(Fig.\ \ref{proc1}(2)) the noise {\it sigma4} filter was applied.
 
\begin{figure}
\centering
\includegraphics[scale=0.6]{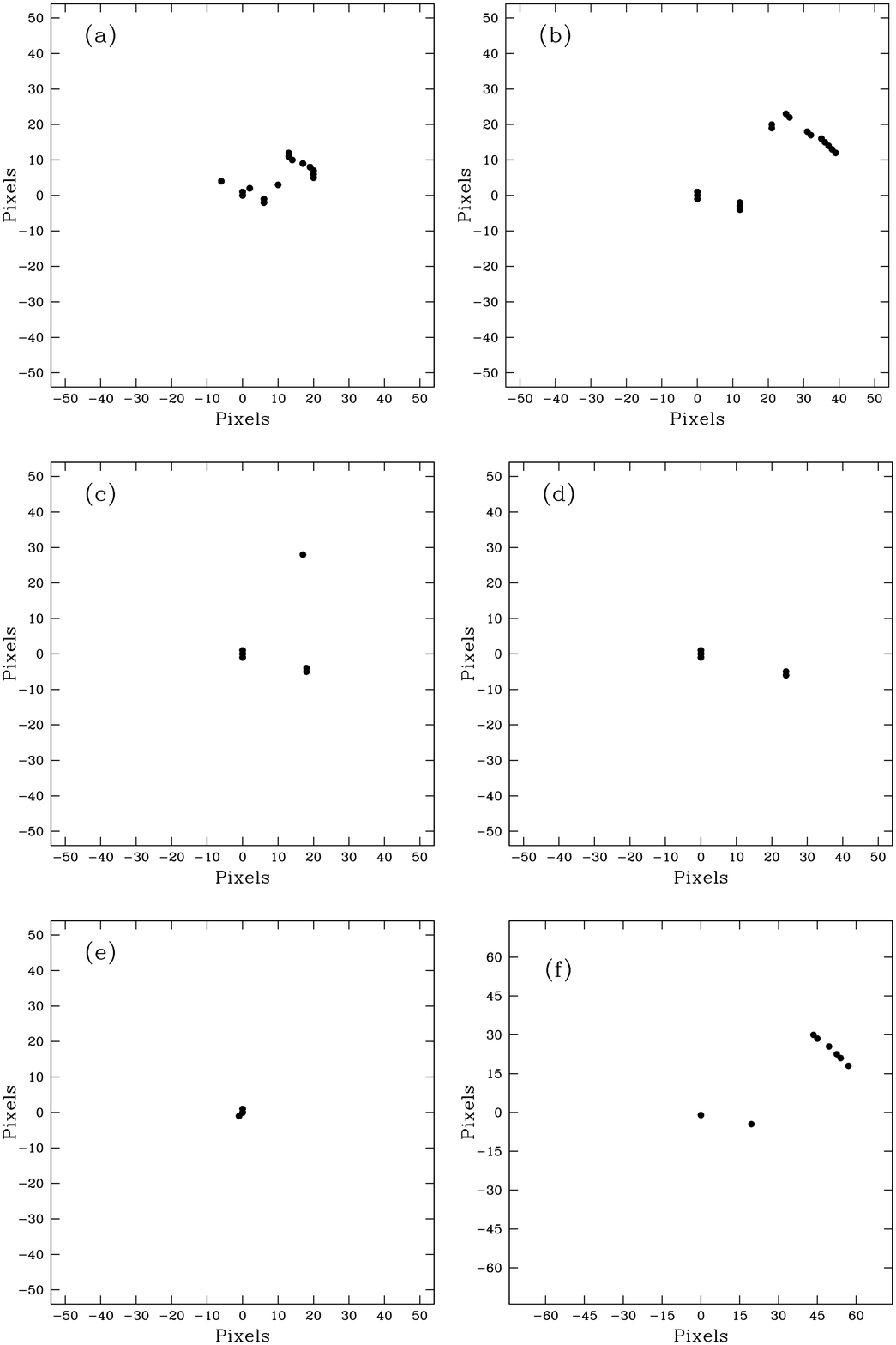}
 \caption{Location of triplets identified in the wavelet spectrum of (a) cc1 ;
 (b) cc2; (c) cc3; (d) cc4; (e) cc5. (f) Location of the validated triplets
 after the comparison of identified triplets for each particular wavelet 
spectrum (as explained in Section \S\ref{we}). Please note that axis scale 
is much larger in this plot to include all the identified layers at
 3$\Delta t$.}
\label{proc2}
\end{figure}

From the wavelet analysis of each cross-correlation after noise
 filtering, we obtain a wavelet spectrum and we analyse the peaks of these
 wavelet spectra as explained in Section \S\ref{we}. For comparison, Figure\ \ref{proc2}
 shows the identified triplets in the wavelet power spectrum corresponding to
 each cross-correlation function analysed on the same scale as the maps in Figure\
 \ref{proc1}. Each point in Figure\ \ref{proc2} corresponds to the position of
 the central peak of each identified triplet. Figure\ \ref{proc2}f shows the
 final triplets selected from the comparison of the validated triplets in
 each cross-correlation. Scale in Figure \ref{proc2}f is larger than 
in the other plots in Figure \ref{proc2} in order to include those triplets that only
 appears in {\it cc1} and {\it cc2}. Bear in mind that the comparison is made at
 3$\Delta t$ (see Section \ref{we}). From the analysis of this particular G-SCIDAR
 measurement we have detected three different turbulence layers moving at
 different velocities. The first identified turbulence layer is the
 dome seeing and  corresponds to the points around the origin of coordinates 
 in Figures \ref{proc1} and \ref{proc2} with velocity equal to zero (the same position at the 
auto-correlation and the five cross-correlations). The second
 turbulence layer is moving from the centre to the bottom right
 of the frames with a relative slow velocity. It corresponds to a low-altitude 
turbulent layer close to or at observatory level. This second layer is
 located at around the pixel (20,$-7$) at 3$\Delta t$ (Figure \ref{proc2}).
 The third turbulence layer identified is moving quickly from the centre
 to the upper right of the frames. This third layer is composed of several
 layers apparently at similar altitudes but moving with different velocities.
 Actually, it seems to correspond to a velocity-stratificated layer formed by
 thinner turbulence layers situated at slightly different altitudes. The wavelet algorithm 
allows us to separate the stratification if the velocity
 difference between the thinner layers is larger than the velocity and altitude
 resolution of the experimental setup.

  The coordinates of the final validated
 triplets are transformed to velocity and altitude using the parameters of the observational
 setup ($\Delta t$,the conjugated plane, observatory altitude, binary star angular separation, etc.). The
velocity and altitude resolution of the proposed algorithm is better than the
current resolution provided by the inversion algorithm for the turbulence
 profile. For this reason, we bin the resolution of the velocity profile to the
 turbulence intensity profile. Such better resolution can be used to obtain the contribution
 of each turbulent layer detected in the cross-correlation to the $C_{N}^{2}$ profile following, for 
example, the approach by Avila et al.\ (2001). We are developing an improved wavelet-based algorithm 
to determine simultaneously the velocity and $C_{N}^{2}$ of individual turbulent layers in order to 
increase the resolution on the determination of the $C_{N}^{2}$ profile. For this subject, we impose the 
condition that the sum of $C_{N}^{2}$ associated with individual layers should be equal to the $C_{N}^{2}$ 
profile derive from the autocorrelation. However, this issue is complex to tackle because of the 
presence of harmonics in the wavelet spectrum and the temporal decorrelation of scintillation (Avila et al.\ 
2001). In this paper, we concentrate on the determination of velocities of turbulent layers and we have 
left the calculation of their associated $C_{N}^{2}$ values for a new and improve version of the current software.

It is important to note that the velocity vector derived from G-SCIDAR 
measurements is in fact the projection of the actual velocity vector of the 
turbulence layer on the plane perpendicular to the observing direction 
(e.g.\ Avila et al. 2001). Such projection affects both the module and
 direction in different ways. The appendix A presents the expresions for
 the errors when deriving the velocity vector from G-SCIDAR observations 
as a function of the actual wind vector of the turbulence layer.

 The developed software does not correct for projection effects on the
 determination of the wind direction because such  projections are small as the
 observations are carried out at zenith angles smaller than 30$^\circ$.
 In particular, errors smaller than four degrees are
 archived when deriving the wind direction (see Appendix A). This is more than
 acceptable taking into account that the wind orientation is not important
 when deriving the relevant temporal parameters for adaptive optics. The
developed software includes the correction of projection effects when
 determining the wind module. The percent errors associated to the wind
 module depend on the wind direction of each turbulence layer respect to 
the azimuth of the observation. For zenith angles of 30$^\circ$, such errors
 could be as larger as 14\% when the actual velocity vector of the turbulence
 layer is on the direction of the azimuth (see Appendix A).

Figure \ref{perfil_030831} shows the
 derived turbulence profile and wind velocity of the particular G-SCIDAR
 measurement analysed in detail in this section.

\begin{figure}
\centering
\includegraphics[scale=0.6]{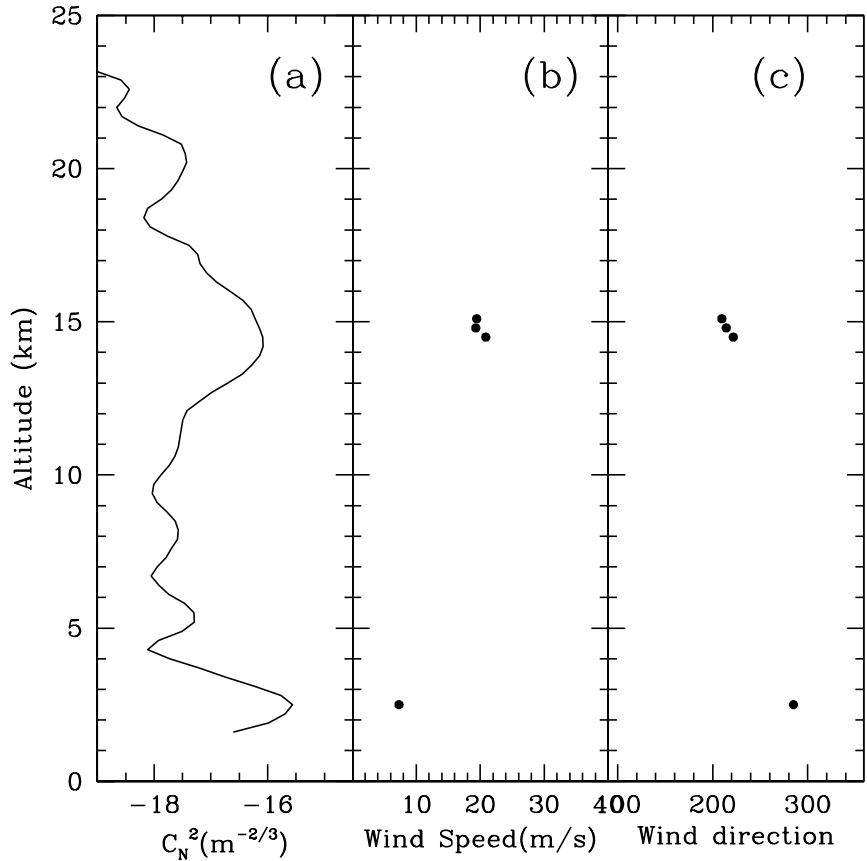}
 \caption{(a) Vertical turbulence profile ($C_{N}^{2}$); (b) wind modulus of
 turbulence layers derived from the wavelet-based algorithm; (c) wind
 direction (in degrees) for the motion of turbulence layers derived from
 the wavelet-based 
algorithm}
\label{perfil_030831}
\end{figure}

\subsection{Comparison of results with balloon measurements}

\begin{figure}
\centering
\includegraphics[scale=0.40]{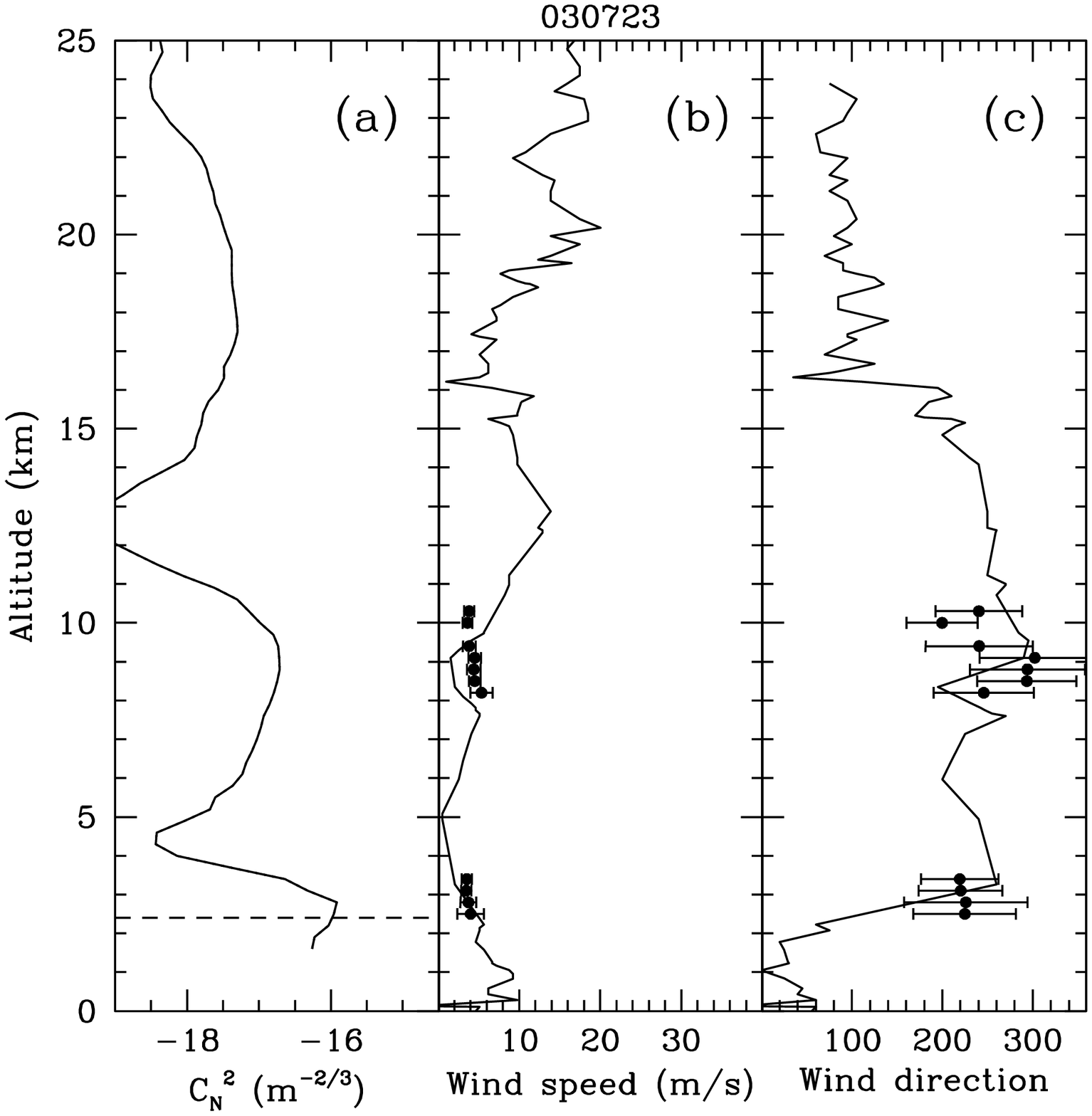}
\includegraphics[scale=0.40]{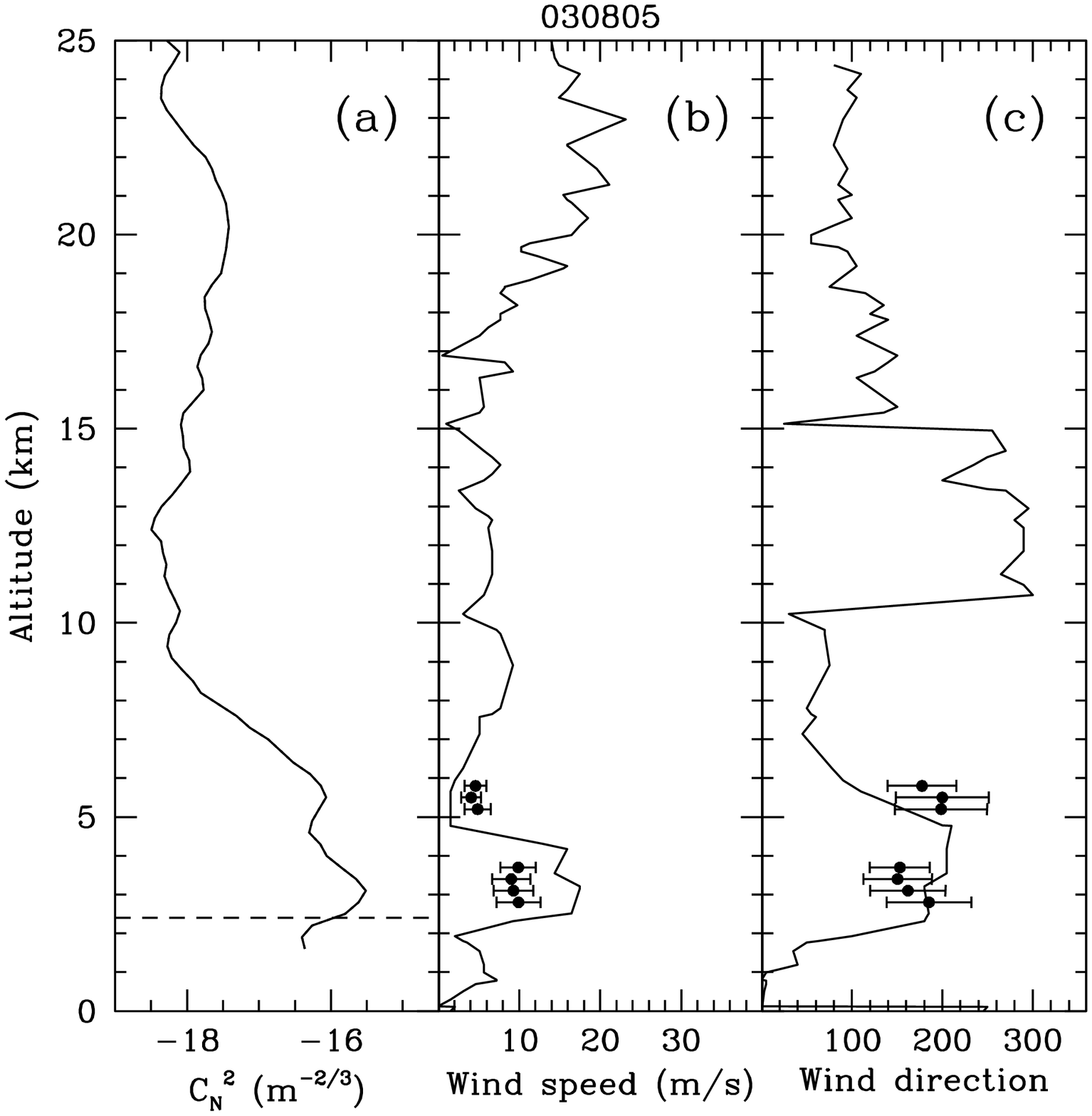}
\includegraphics[scale=0.40]{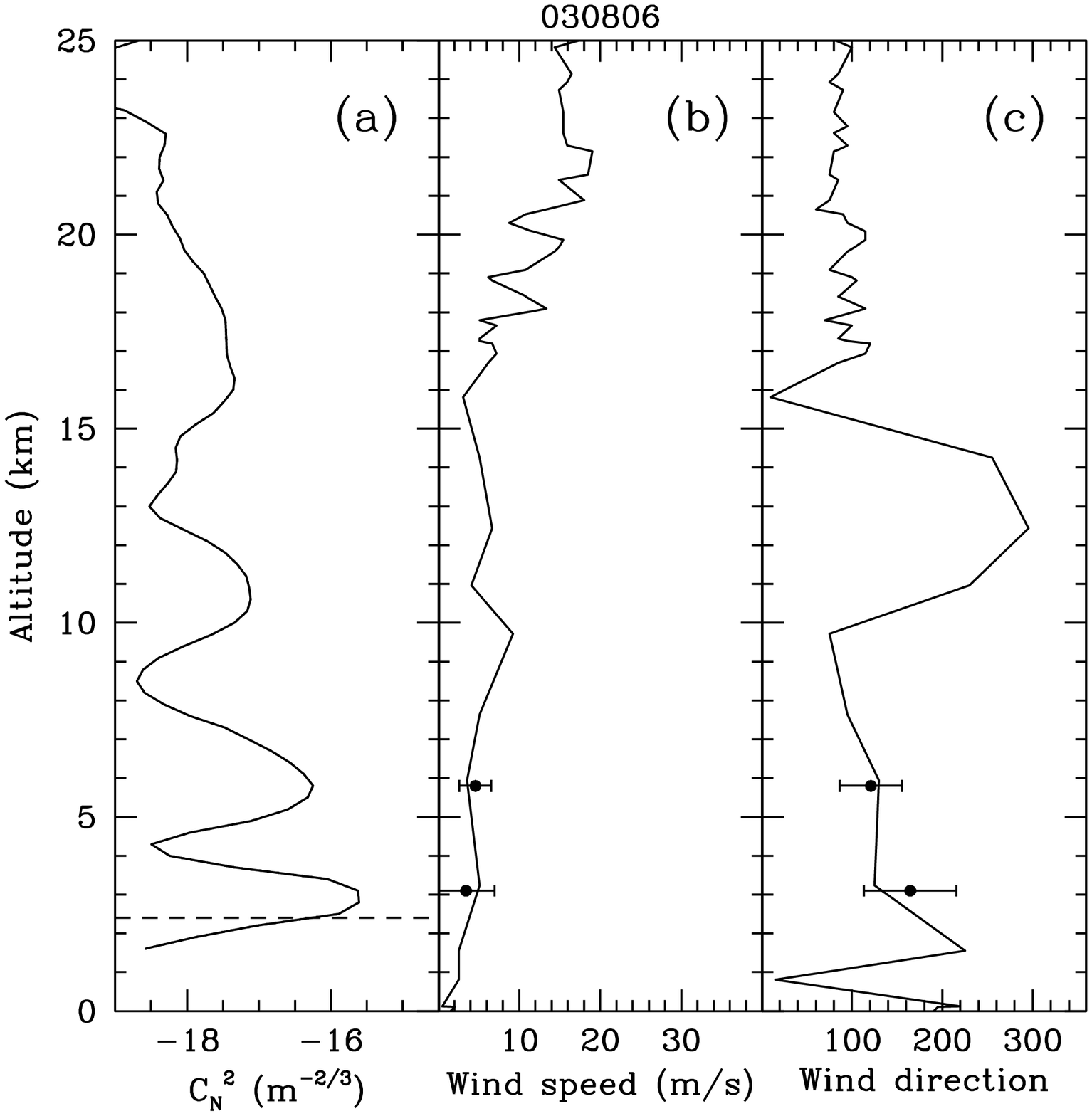}
\includegraphics[scale=0.40]{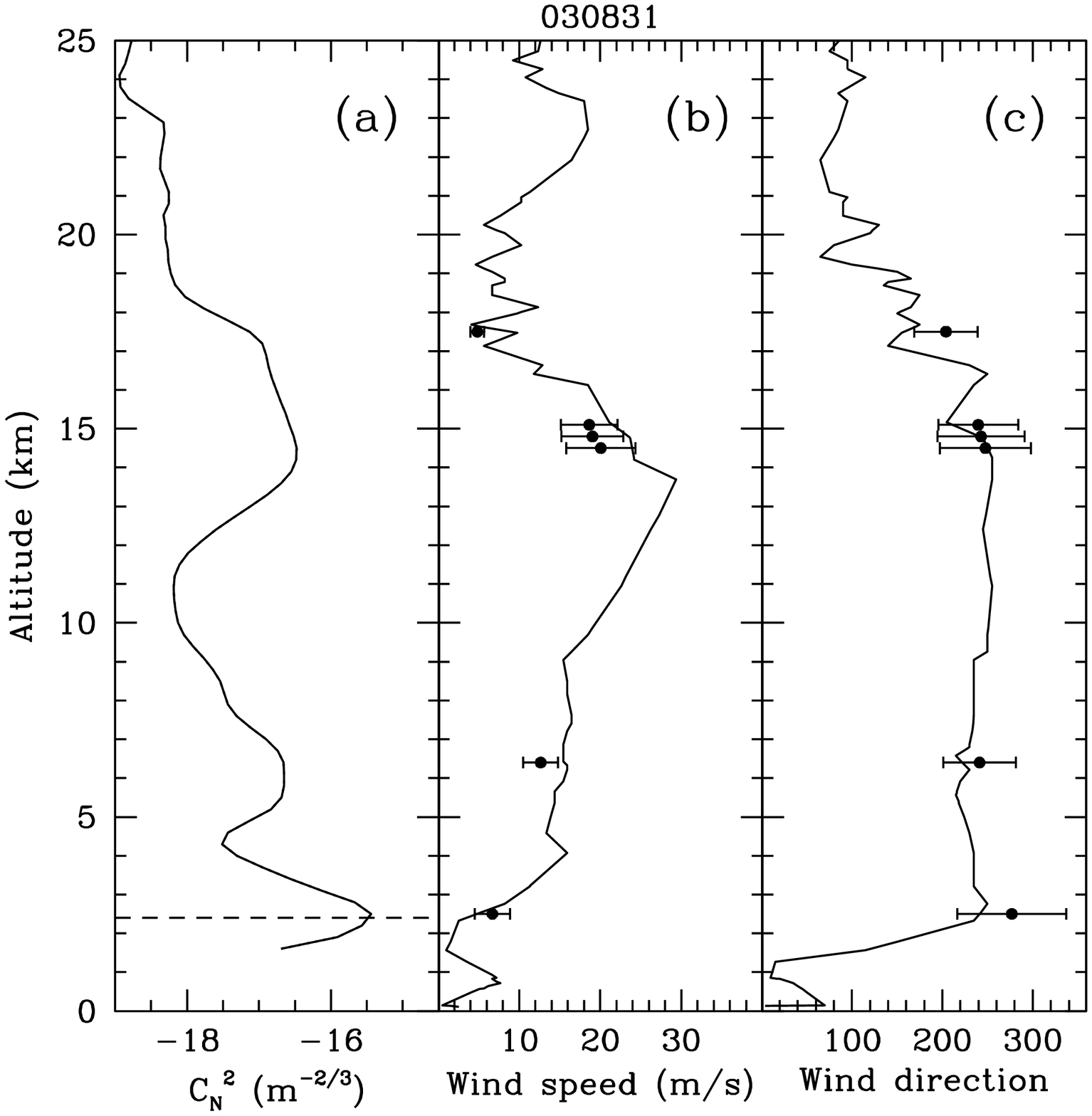}
 \caption{Wind vertical profiles (left) and the simulataneous vertical
 turbulence profile ($C_{N}^{2}$) for four nights (dates are
 indicated at the top of each plot). Error bars only indicates the standard
deviation of the G-SCIDAR measurements (around 130 individual profiles) during the balloon ascent. The solid
 line in the left plot corresponds to wind speed measurements with radiosondes
 launched 13 km away from the Observatory and close to sea level. Filled
 dots are the derived velocities for turbulence layers from the G-SCIDAR
 measurements during the corresponding balloon ascent. The right panel of each
 plot shows the average (solid line) and median (dotted line) of the
 $C_{N}^{2}$ measurements derived from G-SCIDAR observations during the
 corresponding balloon ascent. The horizontal axis represents the $C{_N^2}$ on a
 logarithmic scale. The vertical axis corresponds to the altitude above 
 sea level. The horizontal dashed line indicates the Observatory altitude
 (2400 m). Dome seeing has been removed from the profiles using a procedure 
based on the properties of the parity of functions (Fuensalida, Garc\'{\i}a-Lorenzo \& Hoegemann 2006). }
\label{comparison}
\end{figure}

In order to validate results from the wavelet-based algorithm to determine
 the velocity of turbulence layers, we have used balloon measurements from 
the NOAA station 60018 (see Section \S\ref{data}). We have obtained an average
 $C_{N}^{2}(h)$ and wind profiles from the individual vertical profiles
 obtained approximately during the corresponding balloon ascent. Figure
 \ref{comparison} shows the comparison of both wind measurements, indicating a 
remarkable correspondence between the SCIDAR and balloon measurements. To
 calculate the  wind direction measured with G-SCIDAR we have used the
 position angle in the sky of the binary star observed to derived these
 profiles.

In general, the altitude of turbulence layers in the average $C_{N}^{2}$
 profile appears in regions where the balloon wind data present strong vertical
 gradients. This is the case for three of the four example nights (030723,
 030805 and 030831). This is in good agreement with the idea that wind 
gradients break the stable stratification producing hydrodynamical
 instabilities and generating dynamical turbulence, which leads to the
 formation of turbulence layers in the presence of a potential temperature 
gradient (Coulman et al.\ 1995). The turbulence profile observed for the night 030806
 presents turbulence layers  associated with smooth gradients in
 the wind vertical profile (in modulus or direction) which might 
not be in concordance with
 the accepted scenario. We also identify strong vertical gradients in the vertical 
wind profiles that are not associated with any turbulence layer (night 030805 at around 
10 and 15 km). However, we lack temperature gradient information to analyse those situations properly.

\section{Conclusions}

Monitoring programmes for atmospheric turbulence characterization
are already in progress at several astronomical sites. Such characterization
 will allow us to improve current adaptive optic systems to compensate for
the effects of turbulence on the light from astronomical objects.
The development of powerful mathematical procedures and efficient software 
to treat turbulence data in real time will allow us to develop ``active''
 adaptive optic systems. By ``active'' we mean an adaptive optic
 system using information on the turbulence structure  in real time and
with the capacity of fitting its parameters to the actual turbulence
 conditions.

We have presented a new automatic method for deriving
 the wind 
speed of turbulence layers from G-SCIDAR observations. The developed 
algorithm have a higher capacity of resolving slow turbulence layers
 than previous programs because we make used of five cross-correlation.
 Comparison of 
wind vertical profiles from G-SCIDAR and balloon measurements demonstrates
 the effectiveness of the propose algorithm. We illustrate
 the new algorithm with the results obtained for four nights of G-SCIDAR
 observations at the Observatorio del Teide on the island of Tenerife
 (Spain). The algorithm allows us to process a huge amount of G-SCIDAR
 data without human intervention. The developed software takes into account
 the effect of the projection 
of the actual velocity vector of turbulence layers on the observing direction
 when determining the wind module. The calculation of some important parameters for adaptive optics, such as the coherence time,
can be strongly affected when such effect is not considered.

\section*{Acknowledgments}

The authors thank T. Mahoney for his assistance in editing this paper. We also
 wish to thank A. Eff-Darwich and J. M. Delgado for useful discussions. 
Finally, the author thanks the referee for his constructive comments.

  Wavelet software was provided by C. Torrence and G. Compo, and is available
 at URL: http://paos.colorado.edu/research/wavelets/. This paper is based on
 observations obtained at the Carlos S\'anchez Telescope  at the Teide
 Observatory on the island of Tenerife (Spain). The TCS is a 1.5 m telescope
operated by the Instituto de Astrof\'{\i}sica de Canarias.

 This work was partially funded by the Spanish Ministerio de Ciencia y 
 Tecnolog\'{\i}a (AYA2003-07728).

\section{APPENDIX A: \\ Projection effects on the determination of the velocity vector of turbulence layers from G-SCIDAR observations.}

The velocity derived from Generalized SCIDAR (G-SCIDAR
 hereafter) observations is in fact the projection of the layer velocity 
vector on a plane  perpendicular to the observing direction.
In this appendix, we calculate the mathematical expression for the error in the
 determination of the wind direction from G-SCIDAR due to such projection and assuming turbulence layer moving horizontally..

Let $XYZ$ be a three-dimensional coordinate system. The axes are depicted in a
 world-coordinates orientation with the $Z$-axis pointing up. For an astronomical
 site, the $Z$-axis represents the zenith direction. 

\begin{figure}
   \begin{center}
   \begin{tabular}{c}
  \includegraphics[scale=0.35]{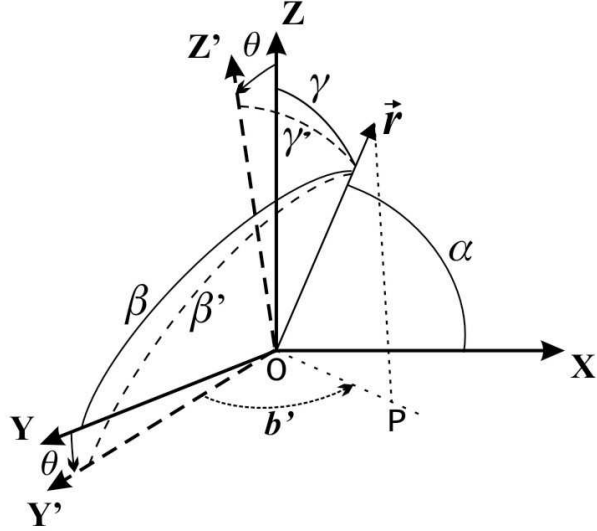}
   \end{tabular}
   \end{center}
   \caption{A view of the coordinate systems $XYZ$ and $XY'Z'$.}
   \label{coor_sys}
   \end{figure}

If we are observing a particular source at a Zenithal distance $\theta$, we 
can define a second three-dimensional coordinate system, $X'Y'Z'$. Such a
 coordinate system can be selected as a  coordinate system rotated about the
 $X$-axis; that is, $X'Y'Z' \equiv XY'Z'$. In this way, the $Z'$-axis represents the
 observing direction, so that, the $Z'$-axis forms an angle $\theta$ with the $Z$-axis, and the $Y'$-axis corresponds to the azimuth direction
 of the observation. Figure
 \ref{coor_sys} shows the $XYZ$ and $XY'Z'$ coordinate systems.

Let (l$_{y}$, m$_{y}$, n$_{y}$) and (l$_{z}$, m$_{z}$, n$_{z}$) be the
 direction cosines of the $Y'$ and $Z'$ axis, respectively, in
 the $XY'Z'$ system expressed in function of $X$, $Y$, and $Z$ in the $XYZ$-system. Then,

\begin{equation}
y' = l_y x + m_y y + n_y z \hspace{1cm} \Longrightarrow \hspace{1cm} y' = 
\cos \theta y  - \sin \theta z 
\label{1}
\end{equation}
\begin{equation}
z' = l_z x + m_z y + n_z z  \hspace{1cm} \Longrightarrow  \hspace{1cm} z' 
= \sin \theta y + \cos \theta z 
\label{2}
\end{equation}

The direction of a vector \v{r} is determined by its direction cosines, 
given by (cos $\alpha$, cos $\beta$, cos $\gamma$) in the $XYZ$-system, and 
(cos $\alpha$, cos $\beta$', cos $\gamma$') in the $XY'Z'$-system (see Fig. \ref{coor_sys}:
\begin{equation}
x_r = |\v{r}| \cos \alpha ; \hspace{0.8cm} y_r = |\v{r}| \cos \beta ; \hspace{0.8cm} z_r = |\v{r}| \cos \gamma 
\label{3}
\end{equation}
\begin{equation}
x' = |\v{r}| \cos \alpha';   \hspace{0.7cm} y' = |\v{r}| \cos \beta' ;   \hspace{0.7cm} z' = |\v{r}| \cos \gamma'
\label{4}
\end{equation}

From equations (\ref{1}), (\ref{2}), (\ref{3}), and (\ref{4}), we can obtain:
\begin{equation}
\cos \beta' = \cos \theta \cos \beta  - \sin \theta \cos \gamma
\label{5}
\end{equation}
\begin{equation}
\cos \gamma' = \sin \theta \cos \beta + \cos \theta \cos \gamma
\label{6}
\end{equation}

The projection of the vector \v{r} on the $XY'$-plane (OP segment in Fig.
 \ref{coor_sys}), the vector \v{r} itself and the $Y'$-axis form a spherical
 triangle (see Fig.\ \ref{coor_sys}). Applying the law of cosines
 (Zwillinger 1995) to this spherical triangle, we can write:
\begin{equation}
\cos \beta' = \cos b' sin \gamma'
\label{7}
\end{equation}
 Replacing equation (\ref{7}) in equation (\ref{5}), we obtain:
\begin{equation}
\cos b' = \frac{\cos \beta \cos \theta - \cos \gamma \sin \theta}{\sqrt{1-(\cos \beta \sin\theta + 
\cos \gamma \cos \theta)^2}}
\label{8}
\end{equation}

For a plane--parallel stratified atmosphere, any turbulence layer will be in the $XY$-plane, and therefore, 
$\gamma=90^\circ$.
 In this case, equation (\ref{8})
can be written as:
\begin{equation}
\cos b' = \frac{\cos \beta \cos \theta}{\sqrt{1-\cos^2 \beta \sin^2 \theta}},
\label{9}
\end{equation}
where $\beta$ corresponds to the direction of the turbulence layer motion referred to the azimuth direction. If a binary star is observed in the direction Z' using the G-SCIDAR technique, the detected direction is the projection in the XY' plane fixed by the b' angle. Therefore, the error is the difference
between $\beta$ and $b'$ and it is given by:

\begin{figure}
\centering
\includegraphics[scale=0.4]{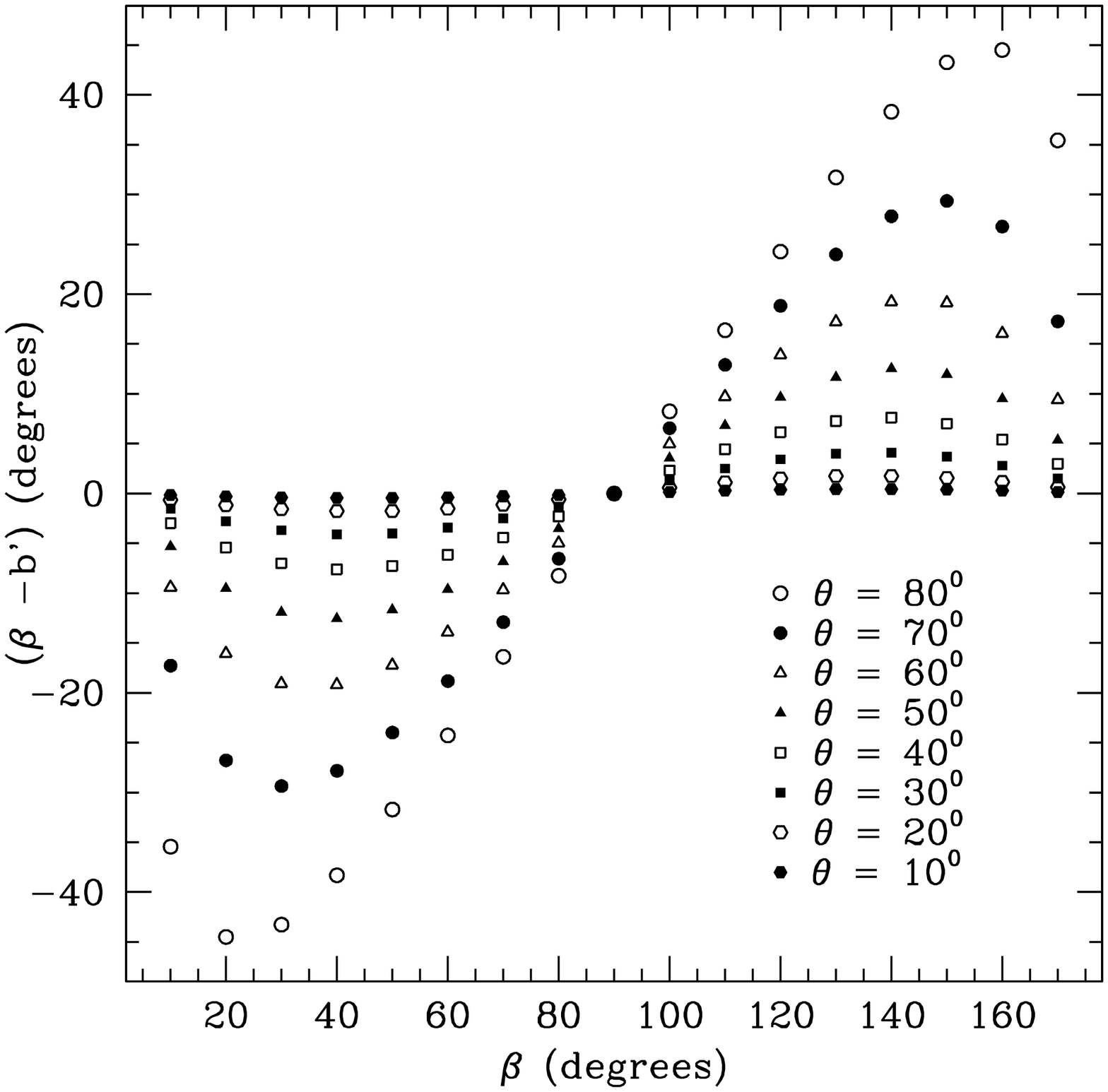}
\includegraphics[scale=0.4]{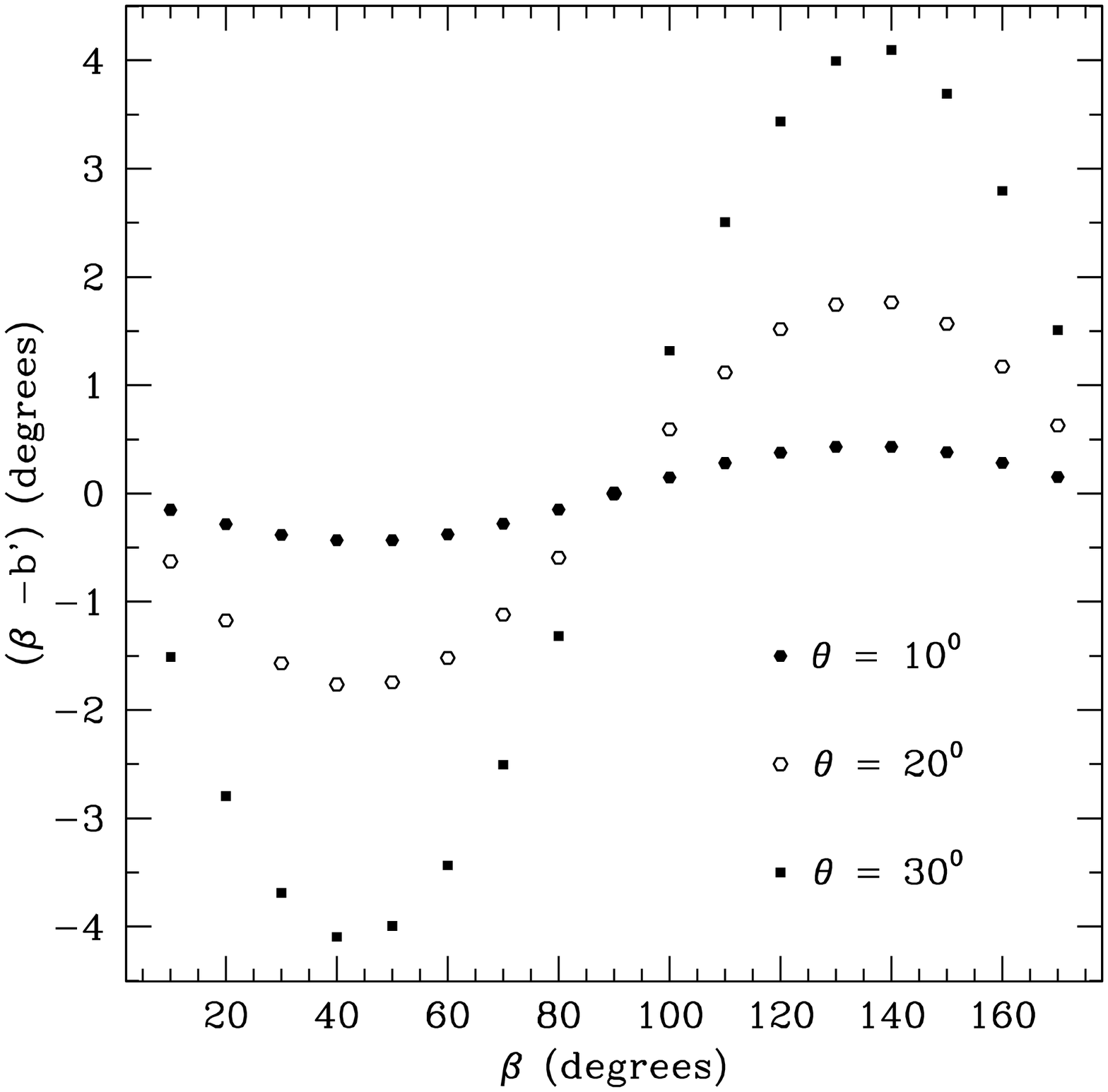}
 \caption{(a) Error in the determination of wind direction from G-SCIDAR
 observations ($b'$) as a function of the actual direction of the velocity vector
 of a turbulence layer ($\beta$). Different symbols correspond to different observing
 directions (zenith angles). (b) The same as (a) but for zenith angles smaller
 than 30 degrees, which is the usual situation in G-SCIDAR observations.}
\label{error}
\end{figure}

\begin{equation}
(\beta - b')_{rad} = \beta - \arccos\{\frac{\cos \beta \cos \theta}{\sqrt{(1 - \cos^2 \beta \sin^2 \theta)}}\}.
\label{10}
\end{equation}

Figure \ref{error} shows the differences between a turbulence layer velocity
 direction and its wind direction determined from G-SCIDAR observations for
 different zenithal distances ($\theta$). For zenith angles smaller than 30
 degrees, such differences are always smaller than 4 degrees. In general,
 G-SCIDAR observations are carry out at zenith angles smaller than 30$^\circ$ 
(this is always the case for G-SCIDAR instruments installed at the Canary
 Islands observatories). Therefore, the effect of the projection of the 
actual velocity vector on the plane perpendicular to the observing direction 
is almost negligible and  is always smaller than 4 degrees.

The projection of the velocity vector on the plane perpendicular to the observing direction also affects to the determination of the velocity module. The proyection of $|$\v{r}$|$$\equiv$r on the XY'-plane is the segment OP$\equiv$r' (see Fig. \ref{coor_sys}), then:

\begin{equation}
r' = r sin \gamma'
\end{equation}

Using equations (9), (11) and (13) and taking into account that we assume $\gamma=90^\circ$ for G-SCIDAR measurements, we derive:

\begin{equation}
\frac{r'}{r}=\sqrt{1-cos^2 \beta sin^2 \theta}
\end{equation}

\begin{figure}
\centering
\includegraphics[scale=0.4]{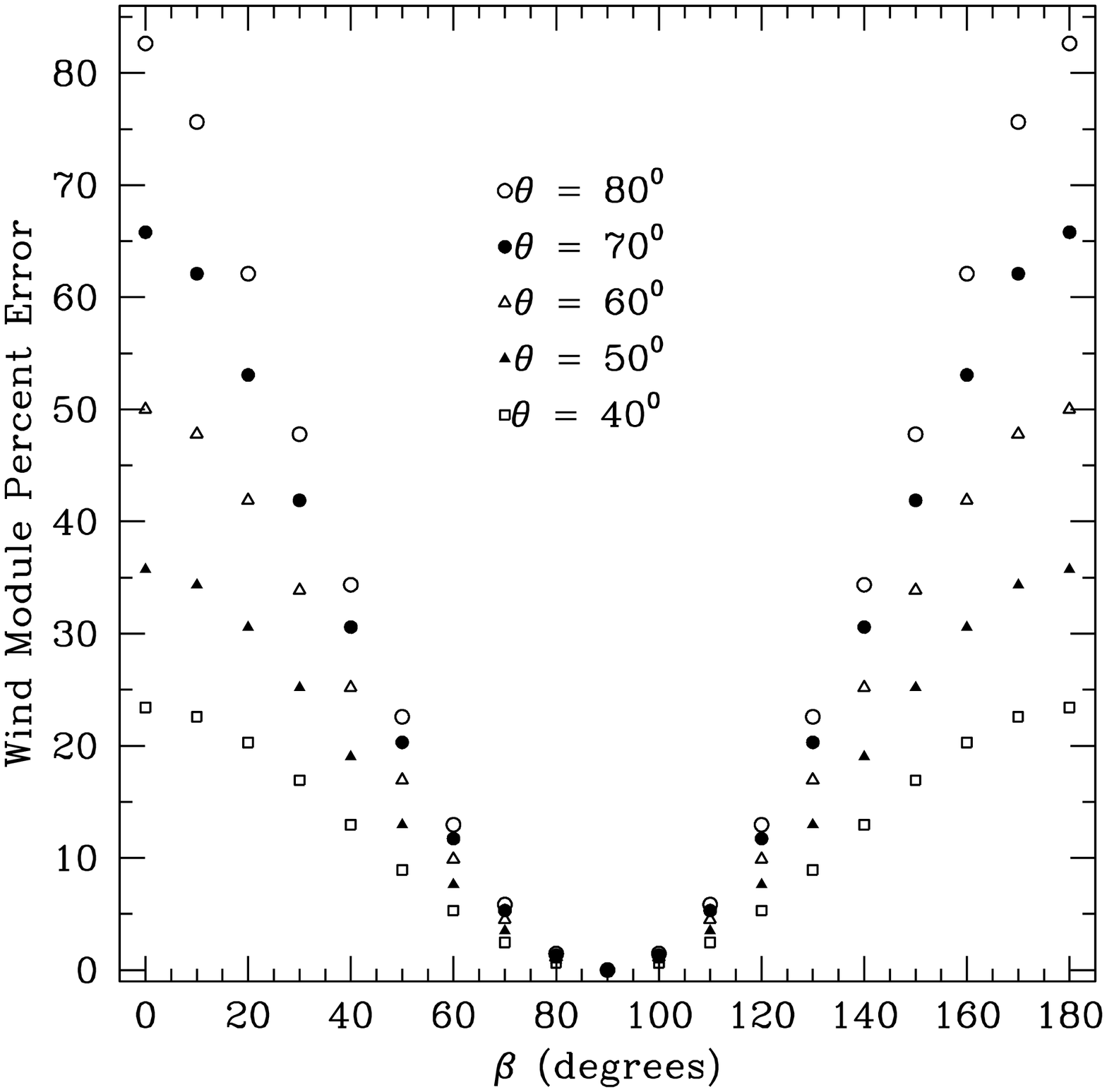}
\includegraphics[scale=0.4]{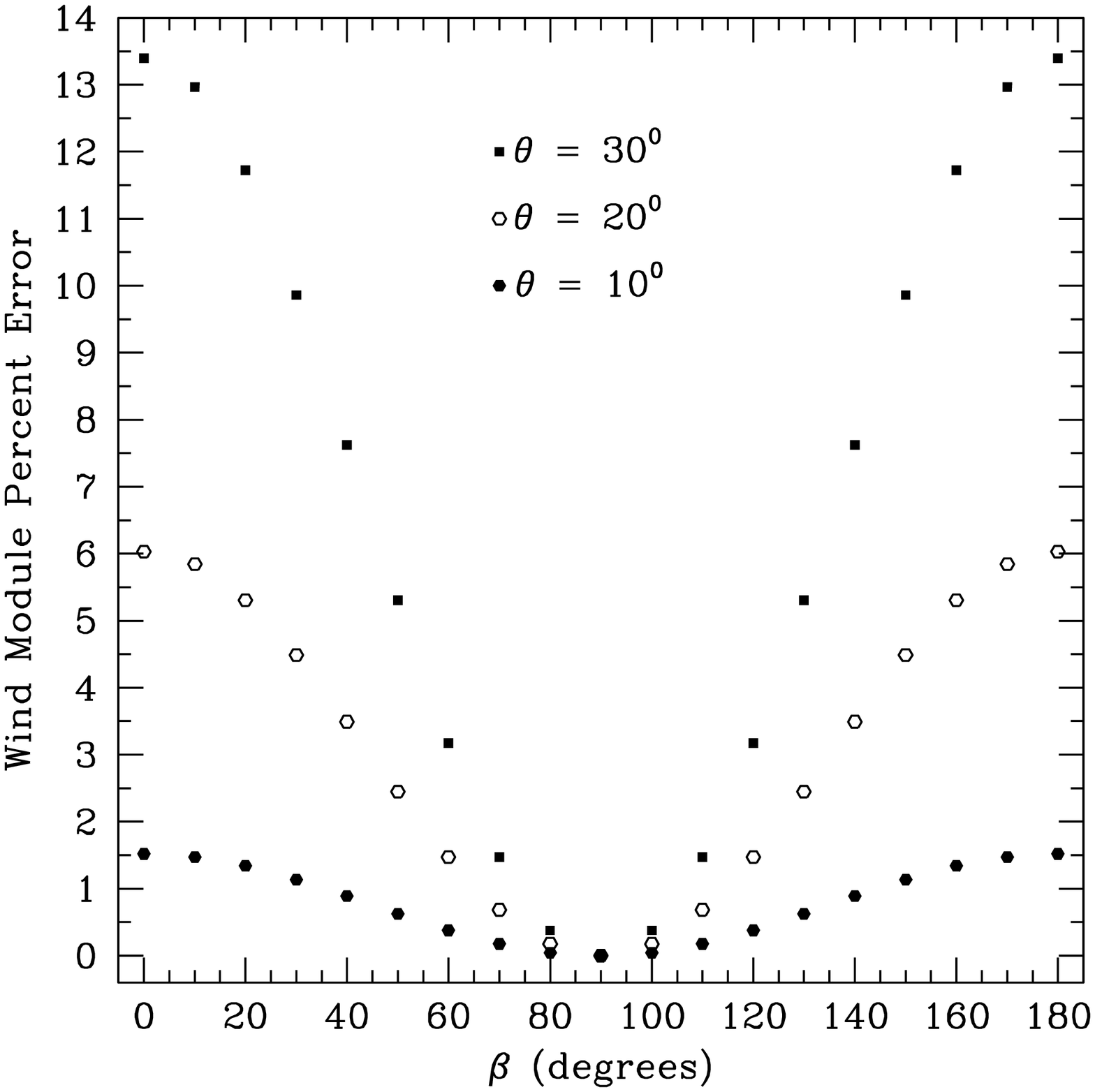}
 \caption{(a) Percent error in the determination of wind modules from G-SCIDAR
 observations as a function of the actual direction of the velocity vector
 of a turbulence layer ($\beta$). Different symbols correspond to different observing
 directions (zenith angles). (b) The same as (a) but for zenith angles smaller
 than 30 degrees, which is the usual situation for G-SCIDAR observations.}
\label{errorm}
\end{figure}

Figure \ref{errorm} shows the percent errors when determining the wind module from
 G-SCIDAR measurements for different Zenith angles. The maximum error  is reached when the velocity vector is along the azimuth direction, $\beta=0^\circ$ or $\beta=180^\circ$. Percent errors smaller than 14\% (see Fig. \ref{errorm}) are archived when observing with zenith angles smaller than 30$^\circ$, which is the usual situation in G-SCIDAR observations.

\end{document}